\documentstyle[aps,epsf]{revtex}  % DO NOT DELETE THIS LINE 
\begin{document}        %  DO NOT DELETE OR CHANGE THIS LINE
\baselineskip 14pt
\title{Studies of $B^0$ Decays for Measuring $\sin(2\beta)$}
\author{Sacha E. Kopp}
\address{Department of Physics, Syracuse University, Syracuse, 
NY 13244 USA }
\maketitle              % Creates the title area, Do Not Remove

\begin{abstract}        % Do Not Delete this line
	Future asymmetric $e^{+}e^{-} \rightarrow \Upsilon$(4S) $B$ 
factories are being constructed to search for CP 
violation in $B^0$ decays.  It is hoped that a CP asymmetry may 
be observed in $B^0\mbox{--}\overline{B}^0$ mixing in analogy with that 
found in neutral kaons.  In this mixing measurement, $B^0$ decays to CP 
eigenstates, such as the rare decay $B^0 \rightarrow \psi K_S$, may 
provide information on the CKM angle $\sin(2\beta)$.  I discuss decay 
additional CP eigenstate branching ratios measured by the CLEO 
experiment that may be used by future $B$ factories to measure 
$\sin(2\beta)$.
\
\end{abstract}   	% Do Not Delete this line

\section{Introduction}               % Introduction goes below.
\vspace{-.15cm}

An important part of the $B$ factory program at SLAC and KEK will be to 
search for CP violation in the mixing and decay of neutral $B$ mesons.  
In close analogy with the kaon system, the weak interaction allows 
mixing of the $B^0$ and the $\overline{B}^0$ through a second-order 
$\Delta B=2$ transition.  The CP violation in this picture results 
from the interference between the amplitude for the decay 
$B^0 \rightarrow f$ to some CP eigenstate $f$ and the amplitude for 
mixing to occur first and then the decay, $B^0 \Rightarrow 
\overline{B}^0 \rightarrow f$.  When the two amplitudes have a 
relative weak  phase, CP violation 
results. Such weak phases arise in the Standard Model because of the complex
CKM matrix \cite{CKM}.

In the Wolfenstein parameterization of the CKM matrix\cite{Wolfenstein}, the phase 
of $V_{ub}$ is $\tan^{-1}(\eta/\rho)$ and both the CKM elements $V_{ub}$ and $V_{td}$
are expected to have large phases.  Thus it is hoped CP violation may be observed in 
the $B$ system.  Figure 1 shows schematically the current 
experimental bounds\cite{stone} on the CKM phase $(\rho,\eta)$ from 
$B_d^0$ mixing, from 
measurements of $V_{ub}/V_{cb}$ and from limits on $B_s$ mixing.  Also 
shown is the bound from measurements of $\varepsilon_{\mbox{K}}$ in 
neutral kaons.  Overlaid on the figure is a triangle whose sides are 
related to products of CKM elements that results from the requirement 
that the CKM matrix is unitary~\cite{jarlskog}:
$V_{ub}V^*_{ud}+ V_{cb}V^*_{cd}+ V_{tb}V^*_{td} \approx V_{ub}+ 
\lambda V_{cb}+ V^*_{td} = 0$  The angles $\alpha$, $\beta$, and $\gamma$ 
are all, in principal, measurable from decays of $B$ mesons.

The phenomonon of $B^0$ mixing is described in a similar way to the 
mixing of the neutral kaons:  Initially pure $|B^0 \rangle$ and $|\overline{B}^0 \rangle$
states are written as orthogonal mixtures of heavy and light $B$ states.
Because the heavy and light states each evolve with their own time-
dependences, there is a quantum-mechanical oscillation of an initially 
pure $B^0$ or $\overline{B}^0$. The weak (CKM) phases in the mixing as well as in 
the decay amplitudes cause
a CP asymmetry in the mixing and decay of initially pure neutral 
$B$'s as they time-evolve:
\vspace{-0.1cm}
\begin{equation}
A_f(t) =\frac{\Gamma(B^0_{phys}(t) \rightarrow f)  -  
\Gamma(\overline{B}^0_{phys}(t) \rightarrow f)} {\Gamma(B^0_{phys}(t) 
\rightarrow f)  + 
\Gamma(\overline{B}^0_{phys}(t) \rightarrow f)}
\end{equation}
\vspace{-0.25cm}

In the $B$ system it is possible to cleanly extract the weak phases from such 
asymmetries, in contrast with the situation in kaon mixing.  In the Standard Model
the asymmetry for $B_d^0$ mixing reduces to \cite{quinn,ali}
\vspace{-0.25cm}
\begin{equation}
A_f(t) = - \mbox{Im} \alpha_f \sin(\Delta m_B t) \\
\end{equation}
\vspace{-0.25cm}
where
\vspace{-0.25cm}
\begin{equation}
\alpha_f = \eta_{CP} \left(\frac{q}{p}\right)_B 
\frac{\langle f|H_{weak}|\overline{B}_q^0 \rangle}{\langle f|H_{weak}
|B_q^0 \rangle} \left(\frac{q}{p}\right)_K
\end{equation}
\vspace{-0.05cm}
and $\eta_{CP}=\pm 1$ is the CP sign of the final state $f$.  The factor $(q/p)_B$
describes the weak phases in the $B$ mixing, and the factor 
$(q/p)_K$ appears whenever the final state $f$ has a $K_{\mbox{S}}$ or 
$K_{\mbox{L}}$ to account for kaon mixing in addition to $B$ mixing 
phases.  The parameter $\Delta m_B$ is the $B_q^0-\overline{B}_q^0$ mass 
difference (for either neutral $B_s^0$ or $B_d^0$ mesons).  

For $B_d^0$ mesons, the phase 
$(q/p)_{B_d} = \mbox{arg}(V_{td}/V^{*}_{td})$ from the presence of intermediate top 
quarks in the box diagram describing $B_d-\overline{B}_d$ mixing.\cite{quinn,ali}   The 
phase of the decay amplitudes depends upon the quark to which the $b$ 
decays.  For $b\rightarrow c$ decays, there is no weak phase since $V_{cb}$ is 
almost real.\cite{frank2}  Thus $b \rightarrow c$ decays in $B_d^0$ 
mixing gives us 
\vspace{-.25cm}
\begin{center}
$\alpha_f = \eta_{CP} \mbox{arg} \left( \frac{V_{td}}{V^*_{td}} \right)$,

$A_f (t) = - \eta_{CP} \sin(2 \beta) \sin(\Delta m_B t)$
\end{center}
\vspace{-.25cm}
It should be noted that in extracting the value of $\sin (2\beta)$ from the mixing
asymmetry, there is a four-fold ambiguity in the actual value of $\beta$ that is 
inferred from this measurement.  In principle, present data, as shown in Figure 1, 
along with recent indications from the CDF experiment\cite{lewis}, indicate that 
$\mbox{sign}[\sin(2\beta)]>0$.  It may be further possible\cite{babar} to 
furthermore remove the last two-fold ambiguity, as mentioned in Section VIII.

In this paper I review several branching ratio measurements of $b \rightarrow c$ 
transitions of $B_d^0$ mesons that are 
relevant to future $B$ factories in measuring $\sin(2 \beta)$.  The 
modes discussed are:

$\begin{array}{lll}
B^0 \rightarrow \psi K_{\mbox{S}} & B^0 \rightarrow \psi K_{\mbox{L}} &
B^0 \rightarrow \psi(\mbox{2S}) K_{\mbox{S}}  \\
B^0 \rightarrow \psi \eta & B^0 \rightarrow \chi_{\mbox{c}1} K_{\mbox{S}} &
 B^0 \rightarrow \psi \pi^0 \\
B^0 \rightarrow \psi K^{*0} & B^0 \rightarrow \psi(\mbox{2S}) K^{*0} &
B^0 \rightarrow D^{*+}D^{*-} \\
\end{array}$

The above 
branching ratios were measured by the CLEO experiment running at the 
symmetric $e^{+}e^{-} \rightarrow \Upsilon\mbox{(4S)} \rightarrow 
B\overline{B}$ CESR collider at Cornell Unversity.\cite{urheim}  The 
goal in studying these decays is to see how many events may be used for 
studying CP violation above and beyond the so-called ''gold-plated'' 
mode of $\psi K_{\mbox{S}}$. The last three decay modes are of course not CP 
eigenstates, but may still be used for studying CP violation, as is 
discussed below.

The observation of CP violation in $B^0$ mixing and decay is a powerful 
first step toward proving CP violation is not simply a feature inherent 
to neutral kaons.  The observation of this kind of CP violation 
will be an important confirmation of the CKM model\cite{frank}.

\section{Data Sample}
\vspace{-.15cm}

Most of the studies presented in this paper represent the yield of 6.3 
fb$^{-1}$ of data, which amounts to $\approx$ 6 
million $B\overline{B}$ pairs.  An additional 3 fb$^{-1}$ was taken 60 
MeV below the $\Upsilon$(4S) to study backgrounds from continuum 
$e^{+}e^{-} \rightarrow q \overline{q}$ light quark production.  

At the $\Upsilon$(4S), it is convenient to use two kinematic 
constraints in reconstructing $B$ mesons from their daughter particles.  
Noting that the $B$ energy is exactly the $e^{\pm}$ beam energy, we 
form the invariant mass of $B$ candidates from $m_B^2 = 
\sqrt{E_{beam}^2 - (\Sigma {\bf p}_{i}^2)}$ since the beam energy is very 
well measured.  Furthermore, we calculate the total energy $E_B$ of our 
$B$ candidates and form the variable $\Delta E = E_B-E_{beam}$ which 
should peak at zero for true signal.  Resolutions on these quantities 
are $\sigma(m_B)\sim 2\mbox{--}3$ MeV and $\sigma(\Delta E)\sim 
10\mbox{--}20$ MeV.

While the analyses which identify final-state $\psi$ mesons differ 
slightly in their selection criteria, all of these studies were done by 
first selecting events where the $\psi$ decays to $ee$ or $\mu\mu$ 
pairs, which comprise just 12\% of the $\psi$ branching ratio.  
For these analyses, lepton selection criteria were imposed for both the 
daughter leptons in the $\psi$ decays, so the efficiencies for 
reconstructing the $\psi$ are typically $\sim 40\%$.  In future $B$ factory experiments 
it may be possible to relax such selection criteria and thereby increase the 
reconstruction efficiencies for these decays.\cite{schuh}  

To suppress the $\psi$'s which come from 
continuum $e^{+}e^{-} \rightarrow c \overline{c}$, a cut of $P_{\psi} < 
2.0 \mbox{GeV}/c$ was imposed.  Thus, unless explictely mentioned 
otherwise, the backgrounds to the modes considered here 
from $B\rightarrow \psi 
X$ decays.  CLEO has previously measured the inclusive $B\rightarrow 
\psi$ momentum distribution\cite{psipaper}, and our Monte Carlo model 
of $B$ decays, which includes several different exclusive $\psi$ modes,
reproduces very well the observed $\psi$ momentum distribution.  
Therefore, many of the background estimates, which 
rely heavily on our Monte Carlo, are well modelled.

\section{$B^0 \rightarrow \psi K_{\mbox{S}}$}
\vspace{-.15cm}

The decay $B^0 \rightarrow \psi K_{\mbox{S}}$ with $K_{\mbox{S}} 
\rightarrow \pi^+\pi^-$ is a so-called ''gold-plated'' CP eigenstate 
because the distinctive signature of two leptons from the $\psi$ and two 
charged tracks emanating from a point detached from the collision point.  
With efficiencies of $\sim 40\%$ for the $\psi$ and $\sim 75\%$ for 
reconstructing the $K_{\mbox{S}} \rightarrow \pi^+\pi^-$, we observe 75 
events in 6.3 fb$^{-1}$, as shown in Figure 2(a).  In 
Figure 2(a), we plot the $\Delta E$ of our $\psi K_{\mbox{S}}$ 
candidates {\it vs.} their mass.  All quantities are plot in units of 
the experimental resolution (see Section 2), so signal 
is expected to lie in a region $\pm 3$ units from the expected values.  
Using our $B^0 \rightarrow \psi X$ MC, we expect $\leq 0.1$ events background 
in the sample.  The branching ratio measured is $\mathcal{B}$$ (B^0 \rightarrow 
\psi K_{\mbox{S}}) = (4.6 \pm 0.06 \; \mbox{stat.} \pm 0.06 \; \mbox{sys.}) 
\times 10^{-4}$, where the first uncertainty is due to statistics and the
second due to systematic uncertainties
(in reconstruction efficiencies and the total number of
$B\overline{B}$ in the sample from which these candidates were selected).
It is hoped we can add to this 75 events using the other decay modes below.

\section{$B^0 \rightarrow \psi K_{\mbox{S}}$, $K_{\mbox{S}} \rightarrow \pi^0\pi^0$}
\vspace{-.15cm}

To identify $K_{\mbox{S}} \rightarrow \pi^0\pi^0$ decays, we search for 
pairs of photons in the CsI calorimeter which are consistent with the 
$\pi^0$ mass using very mass cuts.  When two such pairs are found, 
then the vector defined by the primary collision point and the center-
of-energy of the four photons in the calorimeter is used to define the 
$K_{\mbox{S}}$ flight direction.  The hypothesized $K_{\mbox{S}}$ 
flight distance before decaying is then varied until the two $\pi^0$ 
pairings give the best $\pi^0$ masses. The $K_{\mbox{S}}$ is not used 
in the constraint, but it is found that the $K_{\mbox{S}}$ mass 
resolution improves from $\sim 20$ MeV to $\sim 6$ MeV with this 
procedure.  The $K_{\mbox{S}} \rightarrow \pi^0\pi^0$ reconstruction 
efficiency is 25\%, which, combined with the $\psi$ reconstruction efficiency, 
yields an overall efficiency of about 10\%.  At an asymmetric $B$ factory one
benefits from an additional constraint of matching the $K^0$ origin to
the $B^0$ decay point as measured by the $\psi$ decay.

We observe 
a signal of 15 events, with an expected background of just 0.7 events 
from what is believed to be random photons incorrectly paired with a 
$\psi$ from $B$ decays.  However, we are investigating the possible 
contamination to the sample from $B^0 \rightarrow \psi K^{*}$ decays, 
since these can readily lend some photons and since the $\psi K^{*0}$ 
decay mode has a strong CP component.  This yield, when combined with the 
reconstruction efficiency above, yields a branching ratio of
$\mathcal{B}$$(B^0 \rightarrow \psi K_{\mbox{S}}) = (6.1 \pm 1.6  \; 
\mbox{stat.} \pm 0.13 \; \mbox{sys.}) 
\times 10^{-4}$ (see Table 1), consistent with our
result in Section III.  More importantly, this decay mode adds 15 more events to our
sample of 75 for studying CP violation.

\section{$B^0 \rightarrow \psi K_{\mbox{L}}$}
\vspace{-.15cm}

The $\psi K_{\mbox{L}}$ mode is interesting because in principle it 
presents the same (large) number of events for studying CP asymmetries 
as does $\psi K_{\mbox{S}}$, but the $\psi K_{\mbox{L}}$ has the 
opposite CP.  It should therefore exhibit an asymmetry equal in magnitude but 
opposite in sign as the gold-plated mode.  The difficulty is that the $K_{\mbox{L}}
$ flight path is $\sim 1 \mbox{--} 2$ m, so it doesn't decay within the CLEO 
tracking volume.  It is still plausible to detect the $K_{\mbox{L}}$'s, however, because 
the CLEO CsI calorimeter is 0.81 $\lambda_{\mbox{int}}$ in length, 
hence approximately 65\% of the $K_{\mbox{L}}$ interact 
in the calorimeter and initiate a shower that exceeds 100 MeV in 
energy.  Thus, the signature for the $B^0 \rightarrow \psi 
K_{\mbox{L}}$ decay is the lepton pair from the $\psi$ plus a small 
calorimeter shower from the $K_{\mbox{L}}$.  

The $K_{\mbox{L}}$ shower is a nuclear interaction, is 
quite broad in comparison to showers from $\gamma$'s.  In fact, often 
additional nearby showers are created as 
nuclear fragments travel some distance.  Such shape distinctions are 
used in the selection to successfully reject over 90\% of $\gamma$ 
showers.  In fact, of the 66 $\psi K_{\mbox{L}}$ candidates found, only 
10 have showers that are due to random photons incorrectly paired with 
a $\psi$;  all the rest are real $K_{\mbox{L}}$.  

The $K_{\mbox{L}}$ defines its direction, but not its energy.  We use 
$E_{\psi}$ and ${\bf p}_{\psi}$ along 
with the $K_{\mbox{L}}$ direction to reconstruct $m_B$, but loose the 
$\Delta E$ constraint.  To suppress backgrounds from $B^- \rightarrow 
\psi K^{(*)-}$ we reject events which have an additional track that 
makes a mass $\geq 5$ GeV.  Furthermore, we veto events where the 
$K_{\mbox{L}}$ candidate shower makes the $\pi^0$ mass when paired with 
any other photon in the event in order to reject backgrounds from $B^0 
\rightarrow \psi K_{\mbox{S}}$.

Figure 2(b) shows the results of this search:  66 events are 
found in 6.7 fb$^{-1}$ of data, where 35 are expected to be from 
signal, 31 from backgrounds.  As stated earlier, most of these 
backgrounds are from real $K_{\mbox{L}}$'s from $\chi_{\mbox{c1}} K_{\mbox{L}}$,
$\psi K^{*0}$, and $\psi K^{*+}$.  The CP dilution of these backgrounds is 
15 events.

We have not studied reconstruction efficiencies for this decay mode, although they can 
be inferred from the yield of 35 events in 6.6 fb$^{-1}$ and the known branching ratio 
for $\psi K_{\mbox{S}}$.

\section{$B^0 \rightarrow \psi \pi^{0}$ / $\psi \eta$}
\vspace{-.15cm}

The $\psi \pi^{0}$ mode has the same CP and tests a similar Feynman diagram 
to the often-cited decay mode $B^0 \rightarrow D^{+}D^{-}$.\cite{babar}  
It is color-suppressed relative to $D^+D^-$, but perhaps $\psi \pi^0$ will yield 
more net events because the $D^+D^-$ channel has few subsequent decay modes which
can be reconstructed.  The $\psi \pi^0$ mode would presumably exhibit a 
asymmetry of $+ \sin(2\beta)$.  Because it is the Cabibbo-suppressed 
version of the $\psi K^0$, we can predict the 
branching ratio will be 
$\mathcal{B}$$(B^0 \rightarrow \psi \pi^{0}) = (f_{\pi}/f_K)^2 \tan^2\theta_C$ 
$\mathcal{B}$$(B^0 \rightarrow \psi K^{0}) \sim 6 \times 10^{-5}$ 
or, by isospin conservation, $\mathcal{B}$$(\psi\pi^0) = 0.5\times$$\mathcal{B}$
$(\psi \pi^-) \sim (2.5\pm1.5)\times 10^{-5}$, where I've used the previously 
published CLEO result\cite{psipi} for $\psi \pi^-$.  

We reconstruct $\pi^0 \rightarrow \gamma \gamma$ decays in the CsI calorimeter,
which has an efficiency of $\sim 60\%$.  We observe 7 candidate events with the
background expected to be 0.7 events.  The background comes predominantly 
from real $\psi$'s from $B$ decay paired with random $\gamma$'s in the event
which accidentally form the $\pi^0$ mass.  With a net detection efficiency of 24.3\%, we 
obtain a branching ratio of $\mathcal{B}$$(B^0 \rightarrow \psi \pi^0) = (0.34 \pm 0.16 \; 
\mbox{stat.} \pm 0.04 \; \mbox{sys.})\times 10^{-4}$.  
We used the procedure of Feldman and Cousins \cite{feldman} to obtain the 68\% C.L. 
intervals for the Poisson signal mean.

The $\psi \eta$ decay has the same CP as $\psi\pi^0$, so should exhibit 
the same asymmetry.  By isospin, we might expect that $\mathcal{B}$$(\psi \eta) = 
\frac{1}{3} \times $$\mathcal{B}$$(\psi\pi^0)$.  We searched for this mode using $\eta 
\rightarrow \gamma\gamma$ decays ($BR = 39\%$).  In 6.3 fb$^{-1}$ no 
events were seen.

\section{$B^0 \rightarrow \chi_{\mbox{c1}} K_{\mbox{S}}$}
\vspace{-.15cm}

This final state has the same CP as the $\psi K_{\mbox{S}}$, so should 
have the same sign asymmetry.  We select this decay mode by searching 
for $\chi_{\mbox{c1}} \rightarrow \psi \gamma$, $\psi \rightarrow ll$, 
and $ K_{\mbox{S}} \rightarrow \pi^+\pi^-$ decays. In 6.3 fb$^{-1}$ 6 
events were seen on a background of 0.6 events.  The net detector efficiency
is 16\%, giving a branching ratio of  $\mathcal{B}$$(B^0 \rightarrow \chi_{\mbox{c1}}
K^0) = (4.5^{+2.8}_{-1.8} \pm 0.9) \times 10^{-4}$.  Again, the prescription of 
Feldman and Cousins \cite{feldman} was used.  The background 
is expected to consist of random combinations of $\psi$'s and 
$\gamma$'s.  We are investigating the explicit contribution from $\psi K^*$.

\section{$B^0 \rightarrow \psi(2\mbox{S}) K^{(*)0}$}
\vspace{-.15cm}

The $\psi(2\mbox{S}) K_{\mbox{S}}$ mode is a CP 
eigenstate and the branching ratio is reported here for the first time.  
The $\psi(2\mbox{S}) K^{*0}$ mode has previously been observed by 
CDF\cite{cdfpsiprime} and is not a CP eigenstate:  because the two 
vector particles originate from a spin-0 $B^0$, there is an additional 
factor of $(-1)^l = \pm 1$ in the final state CP due to orbital angular 
momentum between the particles.  Even though the $\psi(2\mbox{S}) 
K^{*0}$ mode is a superposition of two CP eigenstates, it is hoped that 
a single CP state may dominate as with the decay $B^0 \rightarrow \psi 
K^{*0}$ earlier observed by CLEO and CDF.\cite{psikstar}  Even if both 
CP states are prominent, however, Dunietz has suggested that an angular 
analysis may be used to separate the two CP components.\cite{dunietz}  
In this case, one must look at an angular distribution asymmetry that 
develops with proper decay time of the $B^0$ instead of just a decay 
rate asymmetry, so this analysis would require substantially more data.

The $\psi$(2S) is reconstructed through its $l^+l^-$ ($BR = 12\%$) and 
$\psi \pi^+\pi^-$ ($BR = 32\%$) decays, and both $K^{*0} \rightarrow 
K^+\pi^-$ ($BR = 67\%$) and $K^{*0} \rightarrow K_{\mbox{S}}\pi^0$ ($BR 
= 17\%$) decays are considered. In this particular analysis the lepton 
identification was somewhat more stringent than the previous analyses, 
hence the efficiencies are somewhat lower (see Table 1).  
In the case of the $\psi(2\mbox{S}) K^{*0}$ mode, the systematic 
uncertainties are somewhat larger due to our knowledge of the unknown 
helicity amplitudes in this channel (we assume 
$\Gamma_{\mbox{L}}/\Gamma = 0.5$, and theoretically 0.5 - 0.7 is 
expected.  Note that CDF and CLEO measure $\Gamma_{\mbox{L}}/\Gamma = 
0.52 \pm 0.08$ for $\psi K^{*0}$ decays\cite{psikstar}).

A total of 15 $\psi(2\mbox{S}) K_{\mbox{S}}$ and 21 $\psi(2\mbox{S}) 
K^{*0}$ candidates are observed (see Table 1).  The backgrounds, totalling 
$0.4 \pm 0.2$ and $1.9 \pm 0.6$ events, respectively, are predominantly due to 
random combinatorics, but also from $B^- \rightarrow \psi(2\mbox{S}) 
K^{(*)-}$ decays.  The relative yields of different $\psi$(2S) and $K^*$ 
decay modes is consistent with the different detection 
efficiencies and branching ratios.

In addition to being a possible avenue for extracting $\sin(2\beta)$, the 
$\psi(\mbox{2S}) K^{*0}$ mode presented here and the $\psi K^{*0}$ earlier 
measured by CLEO\cite{psikstar} may help resolve two of the four-fold ambiguities
in the extracted value of $\beta$.  The time-development of $B^0\rightarrow 
\psi^{(')}K^{*0}$ decays contains additional terms due to the interference between
the CP=+1 and CP=-1 components.  This additional interference results in additional
terms proportional to $cos(\beta)\sin(\Delta m_B t)$, hence a detailed measurement
of this decay amplitude may eventually resolve the sign of $\cos(\beta)$ and 
remove two of the ambiguities for $\beta$.  Unfortunately, the 
$K^{*0}\rightarrow K_{\mbox{S}}\pi^0$ decay mode has a quarter of the decay rate 
and less than half of the efficiency of the $K^+\pi^-$ mode, so gives a factor 
9 fewer events in our sample; thus, such a resolution will not come concurrently 
with a measurement of $\sin(2 \beta)$.

\section{$B^0 \rightarrow D^{*+}D^{*-}$}
\vspace{-.15cm}

Like the yet unobserved decay mode $B^0 \rightarrow D^+D^-$, this decay 
mode is not expected to be color-suppressed (as is $\psi \pi^0$), but 
it is expected to be Cabibbo-suppressed relative to the more prevalent 
and previously observed decay $B^0 \rightarrow D_{\mbox{s}}^{*+}D^{*-
}$.\cite{dsdpaper}  So, we might expect a branching ratio for 
$D^{*+}D^{*-}$ of approximately 0.1\%.  Like the case of the 
$\psi^{(')}K^{*0}$, this mode 
is mixture of (+) and (-) CP eigenstates, so a mixing 
analysis works if one of the amplitudes dominates or one does a 
time-dependent measurement of an angular asymmetry\cite{dunietz}.

A more complete description of this analysis has recently been 
published.\cite{jaffe}  Here a brief description is given.  To 
reconstruct this mode, both $D^{*+} \rightarrow D^0 \pi^+$ ($BR=67\%$) 
and $D^{*+} \rightarrow D^+ \pi^0$ ($BR = 33\%$) are used, although for 
background reasons only $B^0 \rightarrow (D^0\pi^+)(\overline{D}^0\pi^-
)$ and $B^0 \rightarrow (D^+\pi^-)(\overline{D}^0\pi^-)$ modes were 
considered.  A total of five $D^0$ and six $D^+$ decay modes were 
considered.

To suppress background, two techniques were employed.  The first was 
kinematic, in which a $\chi^2$ variable was constructed to compare the 
reconstructed $D$ and $D^*$ masses for candidates within a given event 
with the known values.\cite{pdg}  The combinations of particles within 
an event with the best $\chi^2$ was chosen, and this combination had to 
have a $\chi^2<20$ (xx\% efficient for signal while xx\% efficient for random 
$B\overline{B}$ and $c\overline{c}$ backgrounds).  The second technique 
was topological, requiring the flight distance between the $D$ and 
$\overline{D}$ decay vertices, as measured by the silicon vertex 
detector, to be inconsistent with zero given our experimental
resolutions.\cite{svx}  This cut was used for events where there was a 
slow $\pi^0$ from the $D^*$ decay.

The results of the search are shown in Figure 3:  4 candidates were
found.  The background, expected to be 0.4 events, is due to 
$e^+e^- \rightarrow c\overline{c}$ continuum and to 
$B\rightarrow D+X$ decays, where a $D$ and a random $\pi$ are 
incorrectly paired to make a $D^*$.  The probability of 0.4 events 
expected background fluctuating to 4 observed events is $1.1 \times 
10^{-4}$.  

The branching ratio is determined to be $(6.2^{+4.0}_{-2.9}\pm1.0) \times 
10^{-4}$, consistent with our expectations and actually quite 
comparable to the $\psi K_{\mbox{S}}$ mode.  Unfortunately, the 
reconstruction efficiency is quite small here, just $\sim  
10^{-3}$, which is in part due to the 8\% from all the branching ratios 
of the $D$'s and $D^*$'s, but also to the 1\% efficiency for 
reconstructing these high multiplicity decays.  We are currently 
investigating the possibility of only partially reconstructing one of 
the $D^*$'s in this decay, and it is furthermore possible that the 
reconstruction efficiency will be higher at an asymmetric $B$ factory 
due to the Lorentz boost of the $B$.  

\section{Summary}
\vspace{-.15cm}

Table 1 shows the yields for the CP eigenstates summarized in this paper.  In 6.6
fb$^{-1}$, 140 events are found amongst all the modes after dilutions are accounted
for.  Thus, in a 30 fb$^{-1}$ dataset which might be accumulated by one of the $B$ 
factories in one year's run, one could anticipate 640 events for studying 
$\sin (2\beta)$.  Furthermore, as discussed in Section 2, some of the selection
criteria in these various searches have not been optimized for CP studies, 
hence the efficiencies might be increased by another factor of $\sim 1.6$, giving
an anticipated number of events (after dilution) to just over 1000 when all 
modes are included.  Furthermore, one can hope for additional background rejection
when working at an asymmetric collider.  Even ignoring such gains,
however, one should be able to measure $\sin(2\beta)$ with an error of $\pm 0.1$.

The fact that the 3-generation Standard Model with a single Higgs multiplet has just one independent CP-violating phase makes all the CP-violating effects all very strongly constrained.  It will be of great interest to see whether the pattern of CP violation in $B$ decays agrees with the prediction of the CKM model, or whether new physics will have to be invoked to understand the (hopefully many!) manifestations of CP violation observed in the next few years.

\section{Acknowledgements}
\vspace{-.15cm}

It is a pleasure to thank my many CLEO colleagues for the opportunity 
to represent them at the Division of Particles and Fields conference.   
The January Los Angeles climate is a pleasant departure from Upstate 
New York.  In particular, I thank Alexey Ershov, Alex Undrus, Andy 
Foland, and David Jaffe, whose work is represented here.  I also 
thank Sheldon Stone and Silvia Schuh for helpful discussions.

\begin{table}
\caption{CP sign, number of events, background, CP dilution (in events), reconstruction 
efficiency (including daughter decay branching ratios), and branching ratio for
the CP eigenstates reported in this paper.  All results are for 6.3 fb$^{-1}$ of data.}
\begin{tabular}{lcccccc}
Decay Mode & CP & \# & Back- &  CP Dilution & Reconstruction  & $\mathcal{B}$ \\
 &  &Events & ground & (evts.) & $\epsilon$(\%) & ($\times 10^{-4}$) \\ \hline
$B^0\rightarrow \psi K_{\mbox{S}}$ & $-1$  & 75 & 0.1 & 0.05 & 30 & $4.6\pm0.8$ \\
$B^0\rightarrow \psi K_{\mbox{S}}$, $ K_{\mbox{S}}\rightarrow \pi^0\pi^0$
& $-1$ & 15 & 0.7 & 0.35 & 9.6 & $6.1 \pm 2.1$ \\
$B^0\rightarrow \psi K_{\mbox{L}}$ & $+1$ & 66 & 31 & 10 &  &  \\
$B^0\rightarrow \psi(\mbox{2S})K_{\mbox{S}}$ \tablenote{This analysis based on 5.6 fb$^{-1}$} & $-1$ & 15 & 0.4 & 0.2 & 0.94 & $5.2 \pm 1.6$ \\
$B^0\rightarrow \chi_{\mbox{c1}}K_{\mbox{S}}$ 
       & $-1$ & 6 & 0.6 & 0.3 & 16.3 & $4.5^{+2.8}_{-1.8}$ \\
$B^0\rightarrow \psi \pi^0$ & $+1$ & 7 & 0.6 & 0.3 & 24.3 & $0.34 \pm 0.16$ \\
$B^0\rightarrow \psi \eta$ & $+1$ & 0 & 0.1 & 0.05 & 15.3 & $< 0.49$ (90\% C.L.) 
\end{tabular}
\end{table}

\vskip -0.5 cm
\begin{figure}[ht]	% in second brace, h=here, t=top, b=bottom	
\centerline{\epsfxsize 2.8 truein \epsfbox{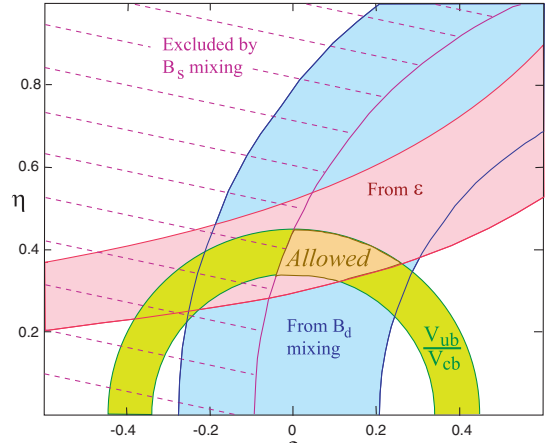}}   
\vskip -.2 cm
\caption[]{
\label{rhoeta}
\small Experimental bounds on the parameters $\rho$ and $\eta$ of 
the CKM matrix.  The allowed region is given by the intersection of the 
bands.  Overlaid is a triangle which results from the requirement that 
the CKM matrix is unitary \cite{stone}.}
\end{figure}

\vskip -0.5 cm
\begin{figure}[ht]	% in second brace, h=here, t=top, b=bottom	
\centerline{\epsfxsize 2.5 truein \epsfbox{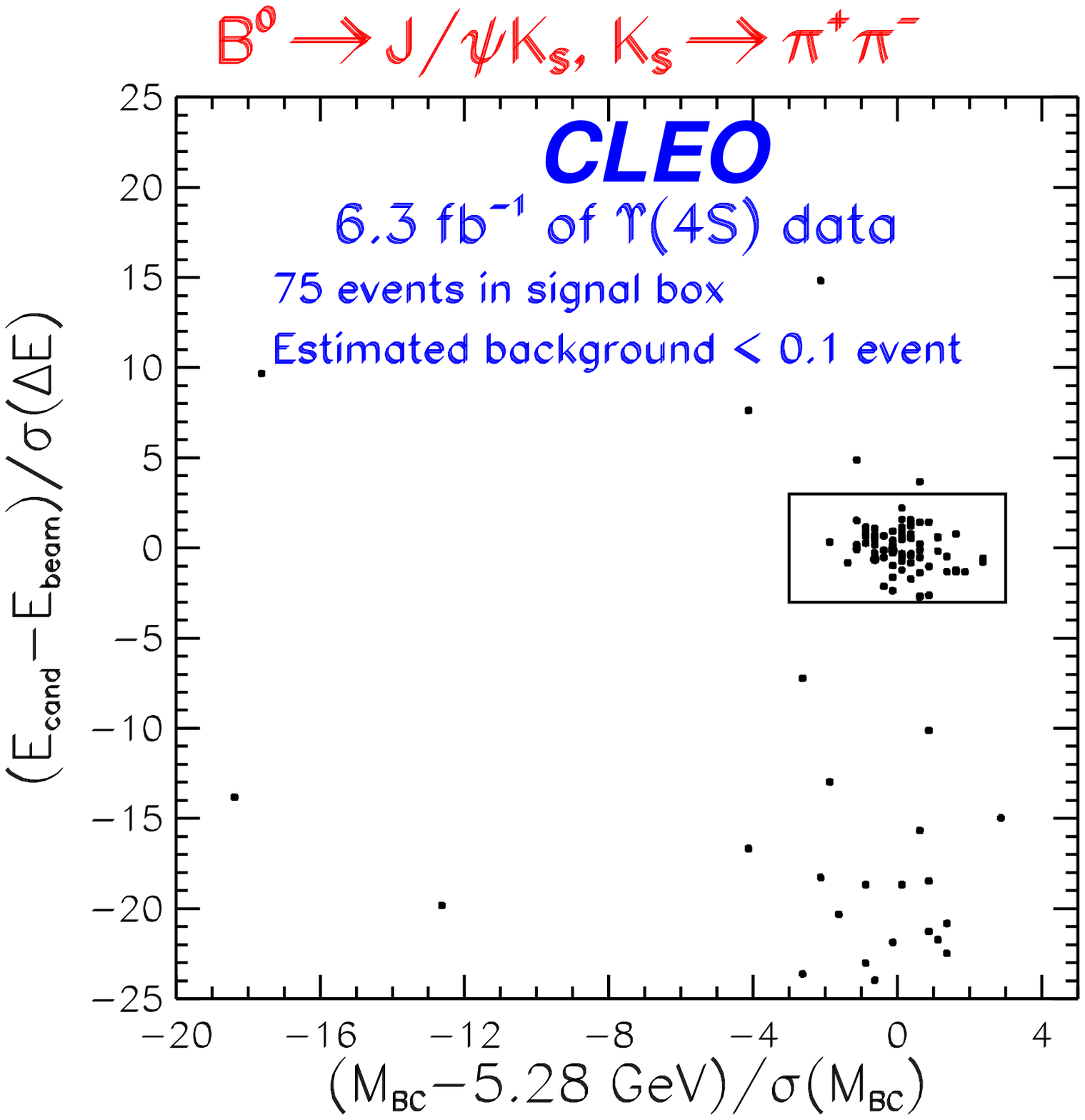}
            \hspace{2.0 cm}\epsfxsize 2.5 truein \epsfbox{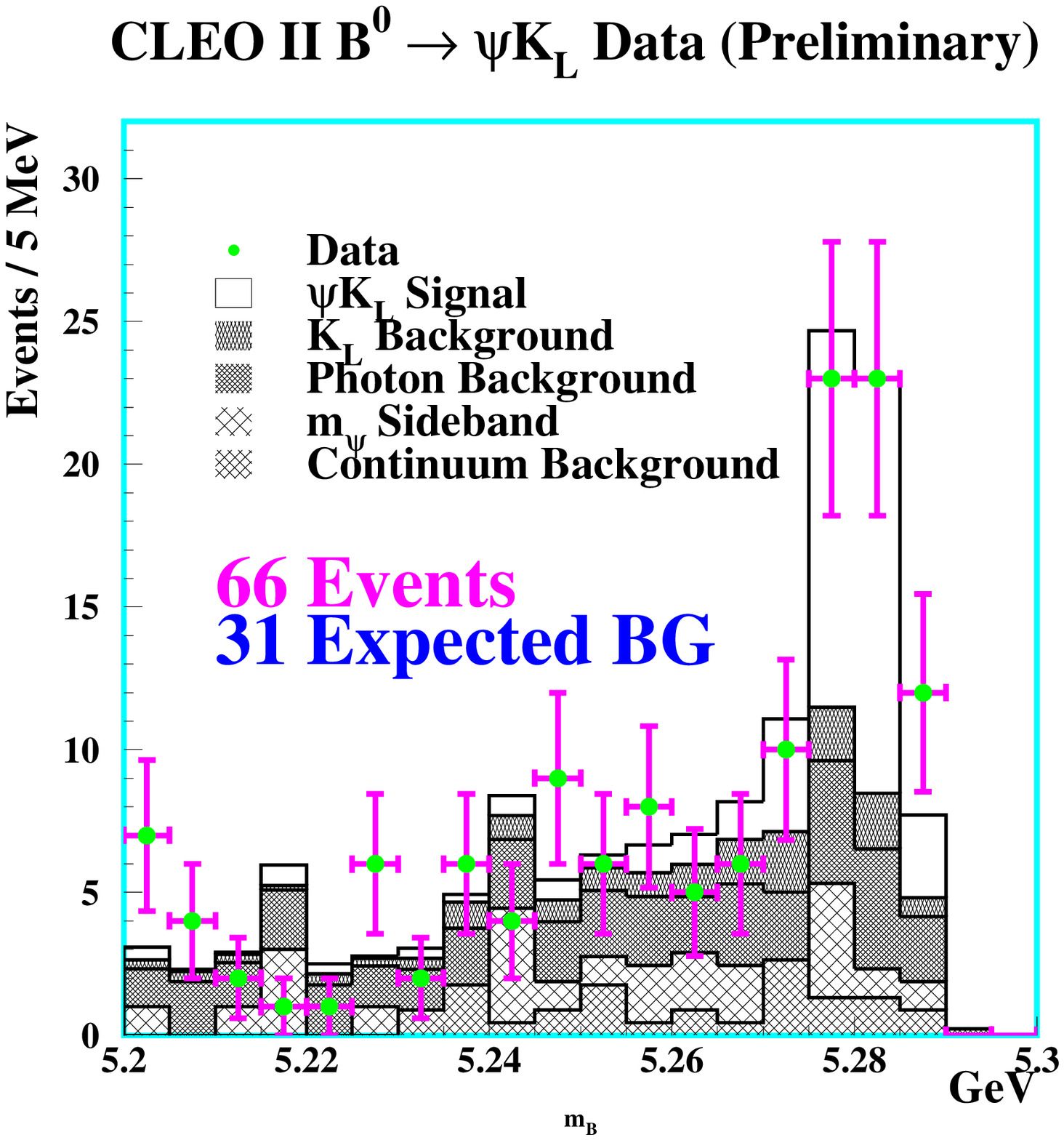}}   
\vskip +0.2 cm
\caption[]{
\label{psik}
\small (a) Reconstructed mass $m_B$ of $B^0 \rightarrow \psi K_{\mbox{S}}$ 
candidates {\it vs.} their $\Delta E$.  (b)  Reconstructed mass for 
$B^0 \rightarrow \psi K_{\mbox{L}}$ candidates, indicating CLEO data (points) and
expectations for signal (open histogram) and various backgrounds (shaded 
histograms).}
\end{figure}

\vskip -0.5 cm
\begin{figure}[ht]	% in second brace, h=here, t=top, b=bottom	
\centerline{\epsfxsize 2.5 truein \epsfbox{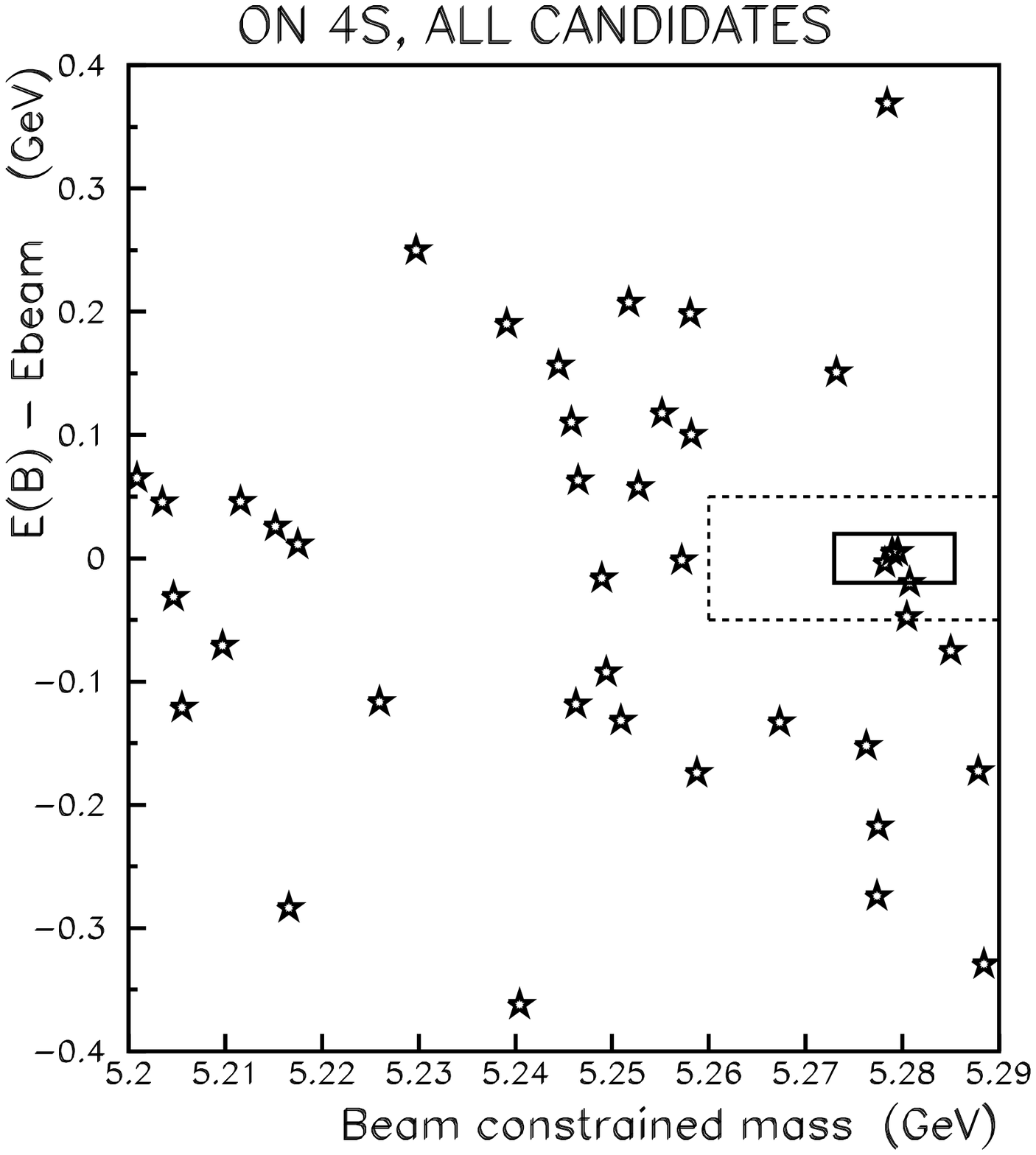}\hspace{1.0cm}
\epsfxsize 2.5 truein\epsfbox{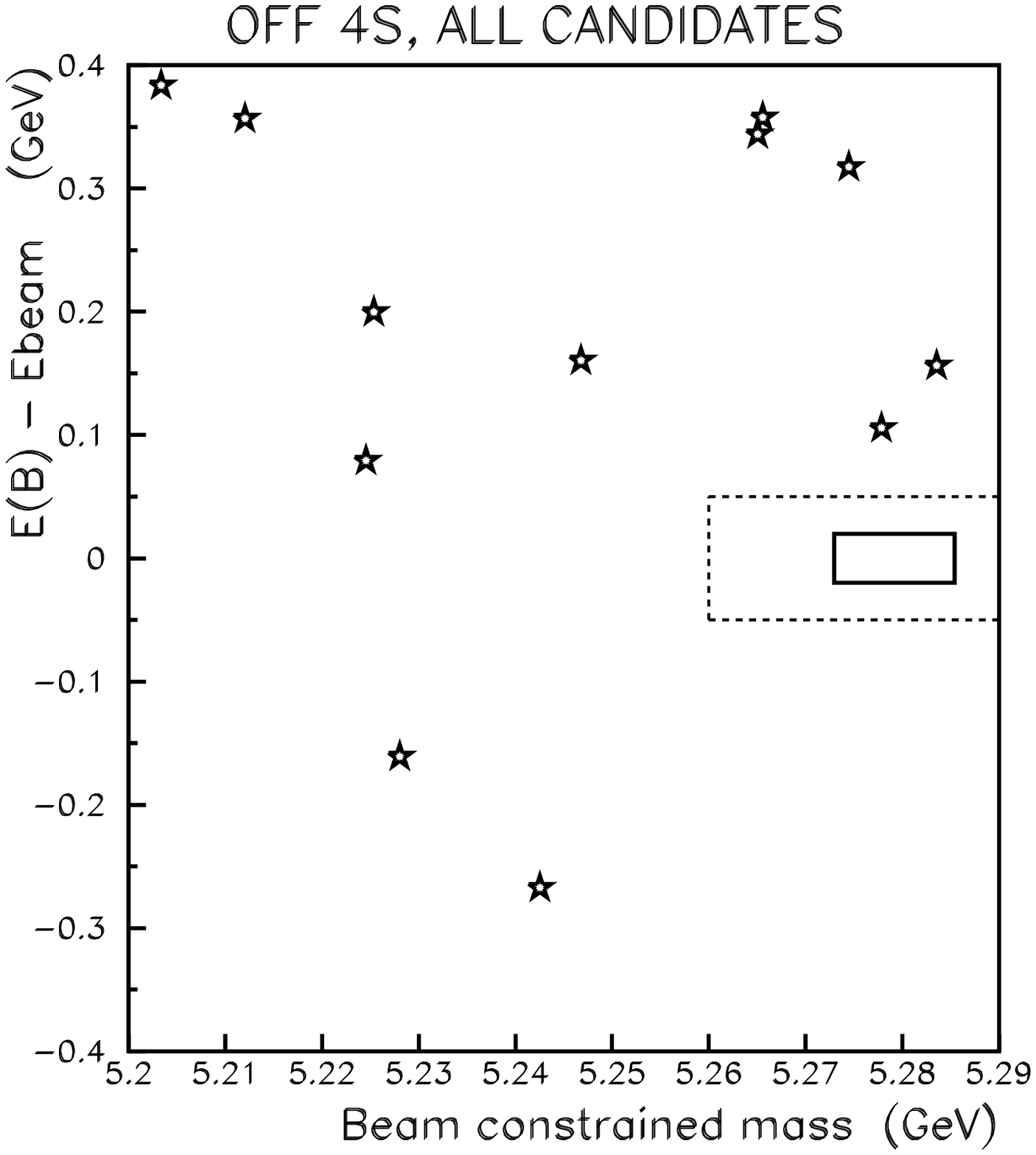}}   
\vskip +.2 cm
\caption[]{
\label{dstarplot}
\small Reconstructed mass $m_B$ of $B^0 \rightarrow D^{*+}D^{*-}$ 
candidates {\it vs.} their $\Delta E$ (a) for data taken at the $\Upsilon(4S)$
resonance;  (b)  for $e^+e^- \rightarrow q\overline{q}$ light quark backgrounds
(including $c\overline{c}$), as studied 60 MeV below the $\Upsilon(4S)$.  }
\end{figure}


\begin{references}

\bibitem{CKM} N.Cabibbo, Phys. Rev. Lett. {\bf 10}, 531 (1963),  
  M.Kobayashi and T.Maskawa, Prog. Theor. Phys. {\bf 49}, 652 (1973).

\bibitem{Wolfenstein} L. Wolfenstein, Phys. Rev. Lett. {\bf 51}, 1945 (1984).

\bibitem{stone}  S. Stone, ``Fundamental Constants from $b$ and $c$ 
Decay,'' Proceedings of Division of Particles and Fields Meeting, 
Albequerque, New Mexico, 1994.

\bibitem{jarlskog}  See the review article by C. Jarlskog, in {\it CP Violation}, C. Jarlskog, ed., World Scientific Press, Singapore (1987).

\bibitem{pdg}  C. Caso {\it et al.}, {\it Eur. Phys. J.} {\bf C31}, 1, (1998)

\bibitem{quinn}  Y.Nir and H.R. Quinn, ''Theory of CP Violation in $B$ 
Decays,'' in {\it $B$ Decays}, ed. S. Stone, World Scientific Press, Singapore 
(1994).

\bibitem{ali}  A. Ali, ''$B$ Decays - Introduction and Overview,'' in 
{\it $B$ Decays}, ed. S. Stone, World Scientific Press, Singapore (1994).

\bibitem{frank2}  For $b\rightarrow u$ decays, there is a phase from  
$V_{ub}/V^*_{ub}$, as discussed in \cite{frank}.

\bibitem{lewis}  See the talk of Dr. Jonathan Lewis at this conference.

\bibitem{babar}  Babar Physics Book, see Chapter 5, ``Determination of $\beta$,'' 
Y. Karyotakis, L. Olivier, J. Smith, and W. Toki, eds.

\bibitem{urheim}  For a description of the CLEO detector, please see 
the summary of the talk given by Dr. John Urheim at this conference.

\bibitem{frank} Note that the observation of CP violation in $B^0$ 
mixing still does not refute the so-called ''superweak'' theory.  To 
truly confirm the KM model, so-called ''direct CP violation'' must be 
observed, such as by looking for decay rate asymmetries like 
$B^{\pm}\rightarrow\pi^{\pm}\pi^{0}$, $B^{\pm}\rightarrow 
K^{\pm}\pi^{0}$, {\it etc}, which was discussed at this conference in 
the talk of Dr. Frank Wuerthwein.

\bibitem{schuh}  For an analysis of  $\psi  K^{(*)}$ decays in which the lepton identification criteria for one of the $\psi$ tracks is relaxed, see the talk by S. Schuh, ``Measurement of the Ratio $f_{\pm}/f_{00}$,'' given at the American Physical Society Centenary Meeting, Atlanta, March 1999. 

\bibitem{psipaper} R. Balest {\it et al.} (CLEO Collaboration),  
{\it Phys. Rev.} {\bf D52}, 2661 (1995).

\bibitem{psipi}  J. Alexander {\it et al.} (CLEO Collaboration) {\it Phys. Lett.}
{\bf B341}, 435 (1995).

\bibitem{feldman}  G.J. Feldman and R.D. Cousins, {\it Phys. Rev.} {\bf D57}, 3873 (1998).

\bibitem{cdfpsiprime} F. Abe {\it et al.}, {\it Phys. Rev.} {\bf D58}
072001 (1998).

\bibitem{psikstar}  C.P. Jessop {\it et al.} (CLEO Collaboration), {\it Phys. Rev.
 Lett.} {\bf 79}, 4533 (1997).  F. Abe  {\it et al.} (CDF Collaboration), 
{\it Phys. Rev. Lett.} {\bf 75}, 3068 (1995), A. Ribonne, talk presented at XIVth 
Rencontres de Physique, La Thuile, Italy (1999).

\bibitem{dunietz}  I. Dunietz, H. Quinn, A. Snyder, W. Toki, and H.J. 
Lipkin, {\it Phys. Rev.} {\bf D 43}, 2192 (1991).

\bibitem{dsdpaper}  D. Gibaut {\it et al.} (CLEO Collaboration), 
{\it Phys. Rev.} {\bf D53}, 4734 (1996); 
H. Albrecht {\it et al.} (ARGUS Collaboration), {\it Z. Phys.} {\bf 54}, 1 (1992).


\bibitem{jaffe}  M. Artuso {\it al.} (CLEO Collaboration), submitted to {\it 
Phys. Rev. Lett.}, hep-ex/9811027.

\bibitem{svx}  The silicon vertex detector was in place for approximately half of the data taken for this 
analysis.  See reference \cite{urheim}.

\end{references}
\end{document}